# Deep Label Fusion: A 3D End-to-End Hybrid Multi-Atlas Segmentation and Deep Learning Pipeline

Long Xie[1], Laura E.M. Wisse[2], Jiancong Wang[1], Sadhana Ravikumar[1], Trevor Glenn[1], Anica Luther[2], Sydney Lim[1], David A. Wolk[3,4], and Paul A. Yushkevich[1]

[1] Penn Image Computing and Science Laboratory (PICSL), Department of Radiology, University of Pennsylvania, Philadelphia, USA
[2] Department of Diagnostic Radiology, Lund University, Lund, Sweden
[3] Penn Memory Center, University of Pennsylvania, Philadelphia, USA
[4] Department of Neurology, University of Pennsylvania, Philadelphia, USA

**Abstract.** Deep learning (DL) is the state-of-the-art methodology in various medical image segmentation tasks. However, it requires relatively large amounts of manually labeled training data, which may be infeasible to generate in some applications. In addition, DL methods have relatively poor generalizability to out-of-sample data. Multi-atlas segmentation (MAS), on the other hand, has promising performance using limited amounts of training data and good generalizability. A hybrid method that integrates the high accuracy of DL and good generalizability of MAS is highly desired and could play an important role in segmentation problems where manually labeled data is hard to generate. Most of the prior work focuses on improving single components of MAS using DL rather than directly optimizing the final segmentation accuracy via an end-to-end pipeline. Only one study explored this idea in binary segmentation of 2D images, but it remains unknown whether it generalizes well to multi-class 3D segmentation problems. In this study, we propose a 3D end-to-end hybrid pipeline, named deep label fusion (DLF), that takes advantage of the strengths of MAS and DL. Experimental results demonstrate that DLF yields significant improvements over conventional label fusion methods and U-Net, a direct DL approach, in the context of segmenting medial temporal lobe subregions using 3T T1-weighted and T2-weighted MRI. Further, when applied to an unseen similar dataset acquired in 7T, DLF maintains its superior performance, which demonstrates its good generalizability.

## 1 Introduction

Deep learning (DL) algorithms generate state-of-the-art performance in segmenting anatomical structures in medical images. However, to reach their full potential, they require relatively large amounts of manually labeled training data, which may not be practical in some applications, such as hippocampal subfield segmentation. In addition, the generalizability of DL methods to data that is not well represented in the training sample is poor. On the other hand, multi-atlas segmentation (MAS) has been shown to generate promising segmentation using relatively small amounts of training data and generalize well on unseen data due to its intrinsic strong spatial constraints and robustness to variability of anatomical structures and image intensity. MAS has two main components: (1) non-linearly registering and warping a set of atlases, i.e. images with the structure of interest manually labeled, to the target image and (2) performing label fusion to derive a consensus segmentation of the target image by combining the warped atlas segmentations (candidate segmentations), typically via weighted voting [1]. Majority voting (MV) [2], the simplest multi-atlas segmentation scheme, gives equal weights to all the atlases at each spatial location. Segmentation accuracy can be improved by assigning spatially varying weights based on local similarity between atlas and target patches, referred to as spatially varying weighted voting (SVWV) [3, 4]. Instead of treating each atlas independently as in SVWV, joint label fusion (JLF) [5] takes correlated errors among atlases into account when estimating weights and yields better label fusion accuracy.

A hybrid algorithm taking advantage of the high accuracy of DL and better generalizability of MAS is desirable, especially in applications that have limited training data. Indeed, there have been a number of attempts to integrate MAS and DL. Some recent studies have used DL to improve the detection of similar (or dissimilar) patches between the target and atlas, for deriving weights in SVWV. Sanroma et al. [6] and Ding et al. [7] uses neural network to learn a nonlinear embedding to transform the patches to a feature space in which the discriminability of the sum of squared difference (SSD) metric is maximized and generates weight maps for SVWV using SSD in this feature space. Ding et al. [8] and Xie et al. [9] directly apply DL to estimate the probability of an atlas having an erroneous vote either at a patch level [9] or the whole image level [8]. SVWV or JLF are then integrated with the probability estimation to generate the final segmentation. A common limitation of the above approaches is that improving the ability to estimate the probability of an atlas having an erroneous vote does not directly translate to the label fusion accuracy. Indeed, as found in [8], a 2% improvement in discriminating erroneous votes translates to only 0.4% improvement in the final segmentation accuracy, similarly seen in [7] and [9]. In the current study, we hypothesize that greater improvement in segmentation accuracy can be realized by incorporating label fusion into an *end-to-end hybrid MAS-DL segmentation pipeline* with a loss function that directly reflects segmentation accuracy. The most relevant prior work is [10], in which authors propose an end-to-end label fusion pipeline consisting of feature extraction and label fusion subnets. Only the feature extraction subnet has learnable parameters, while the label fusion subnet has a fixed structure mimicking the conventional label fusion procedure. Since the label fusion subnet is designed to be differentiable, the network can be trained end-to-end. However, it is only evaluated on binary segmentation of 2D images, which limits its application.

In this work, we propose deep label fusion (DLF, Fig. 1) with the following contributions: (1) This is the first 3D hybrid MAS-DL pipeline that can be trained end-to-end. (2) Promising experimental results show improvements compared to conventional label fusion and direct DL methods in a difficult real-world multi-label problem of segmenting medial temporal lobe (MTL) subregions on multi-modality 3T MRI. (3) When applying to an unseen 7T MRI dataset, we demonstrate that the proposed DLF inherits the good generalizability of MAS. (4) The network is designed to accept a variable number of atlases in testing, which is a unique feature for an end-to-end pipeline.

## 2 Materials

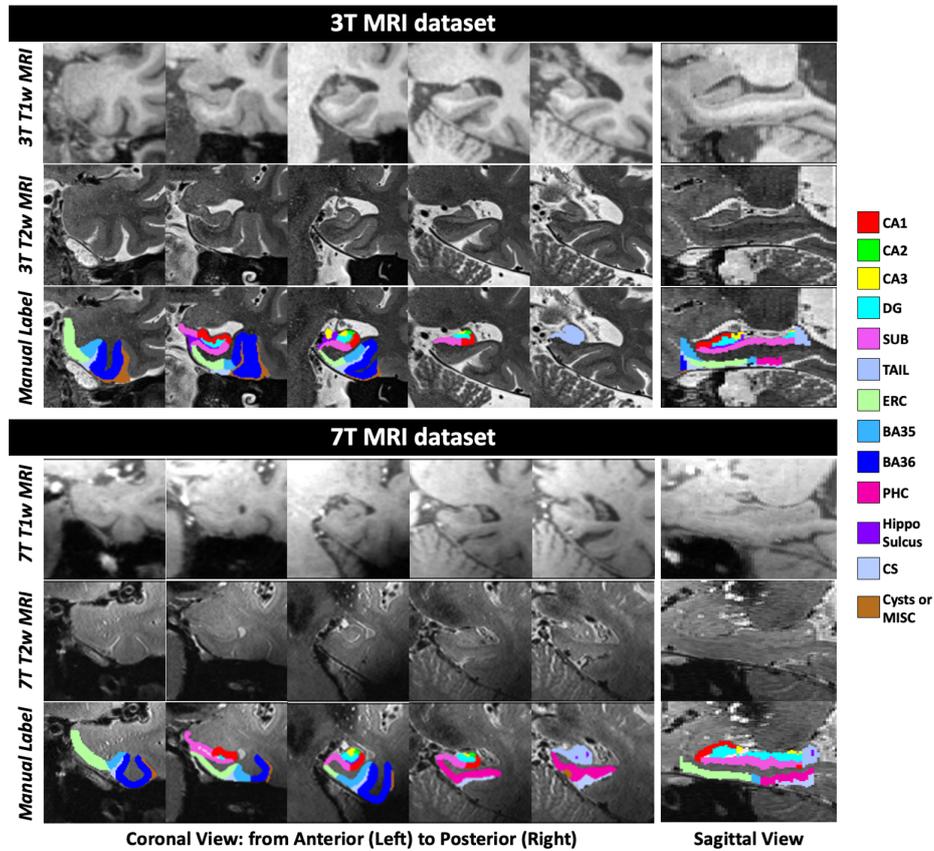

**Figure 1.** Examples T1w and T2w MRI images and manual segmentation of the 3T (top) and 7T datasets (bottom).

A multimodal structural 3T MRI dataset (3T MRI dataset) from XXX was used in this study to develop and validate the proposed DLF algorithm. It consists of T1-weighted (T1w, 0.8x0.8x0.8 mm$^3$) and T2-weighted (T2w, 0.4x0.4x1.2 mm$^3$) MRI scans of 23 subjects [12 with mild cognitive impairment (MCI) and 11 cognitively normal controls (NC)] together with the corresponding manual segmentations of the bilateral MTL sub-regions in the space of the T2w MRI (example in Fig. 1). There are $N_{label} = 15$ labels in total including background, 10 gray matter (GM) labels [cornu ammonis (CA) 1 to 3, dentate gyrus (DG), subiculum (SUB), the tail of hippocampus (TAIL), entorhinal cortex (ERC), Brodmann areas 35 and 36 (BA35/36) and parahippocampal cortex

(PHC)] and 4 supporting non-GM labels [hippocampal sulcus, collateral sulcus (CS), cysts in the hippocampus and miscellaneous (MISC) voxels around the cortex]. Labeling MTL subregions requires exquisite understanding of neuroanatomy, is done in consultation with multiple histological references, and takes many hours per individual. Since the number of available labeled datasets is small, we utilize cross-validation experimental design, with the dataset randomly divided into 4 folds (6 subjects in each of the first 3 folds and 5 subjects in the last fold) keeping the proportion of MCI and NC subjects in each fold similar. A unique aspect of this dataset is that there is no perceivable difference in contrast between most adjacent GM subregions, and boundaries are defined based on anatomical landmarks and geometric rules. Spatial context is required to correctly determine the extent of CA2, CA3, TAIL and the cortex subregions. These factors make the segmentation problem more challenging and well-suited to compare the proposed work to general DL frameworks, such as the U-Net [11].

In addition, a similar MRI dataset (7T MRI dataset) acquired from a 7T rather than 3T MTI was used to serve as an independent out-of-sample dataset to evaluate the generalizability of the proposed approach. Similar to the 3T dataset, the 7T one also consists of T1w (0.7x0.7x0.7 mm$^3$) and T2w (0.4x0.4x1.0 mm$^3$) MRI scans of 24 subjects. The two datasets differ in the tissue contrast and image resolution. In addition, although the manual segmentations were generated according to the same segmentation protocol, subtle difference in placement of tissue boundaries is expected because they are segmented by different persons. Fig. 1 shows examples from the two datasets.

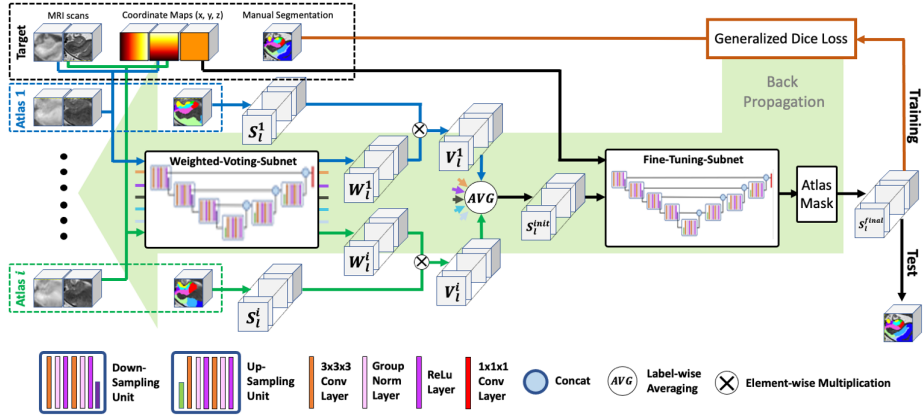

**Figure 2.** Network architecture of the proposed deep label fusion network.

## 3 Method

### 3.1 Deep Label Fusion

Fig. 2 summarizes the two-step structure of the proposed DLF network: (1) *Weighted-voting-subnet*: takes in a pair of target-atlas images to generate label-specific weight maps for each candidate segmentation, i.e. the registered atlas manual segmentation, and performs weighted averaging to produce an initial segmentation and (2) *Fine-tuning-subnet*: corrects errors of the initial segmentation to improve segmentation accuracy. Both subnets adopt a U-Net architecture [11]. Details are described below.

### 3.1.1 Network Architecture

***Weighted-voting-subnet:*** The weighted-voting-subnet takes T1w and T2w target and atlas MRI patches (i.e., 4 channels of dimension $N_x \times N_y \times N_z$) as well as the coordinate-maps with coordinates in the whole target image (3 channels for coordinates in x, y and z directions) as input and generates a 15-channel feature map with the same dimension of $N_x \times N_y \times N_z$, serving as the label-specific weight maps (one for each label). This is different from conventional weighted voting methods, which only compute one weight map for all the labels. The subnet adapts the U-Net architecture that consists of three consecutive down-sampling units (two repetitions of 3x3x3 convolutional, batch normalization [12] and ReLu layers [13], then follow by one 2x2x2 pooling layer) followed by three consecutive up-sampling units (one 3x3x3 transpose convolutional layer with 2x2x2 stripe, two repetitions of 3x3x3 convolutional, batch normalization and ReLu layers). The padding is set to 1x1x1 to preserve the size of the feature maps. The number of feature maps is doubled at each level and the initial number of feature maps is set to 32. Skipped connections are established between each level of up/down sampling. A 1x1x1 convolutional layer is applied in the end to generate the label-specific weight maps. For the $i^{th}$ atlas-target pair ($i = 1, 2, \ldots N_{atlas}$), the weighted-voting-subnet (the same network for all the atlases) is applied to generate label-specific weight maps [denoted as $W^i = \{W^i_l, l = 1,2, \ldots, N_{label}\}$ with voxel value $w^i_{ln}$] for the corresponding candidate segmentation ($S^i$) [which can be converted to a set of binary segmentations $S^i = \{S^i_l, l = 1,2, \ldots, N_{label}\}$ for each label with voxel value $s^i_{ln}$]. Then, elementwise multiplication between $W^i$ and $S^i$ is performed to generate the vote-maps for all the labels [$V^i = \{V^i_l, l = 1,2, \ldots, N_{label}\}$ with voxel value $v^i_{ln}$], i.e. $v^i_{ln} = w^i_{ln} \times s^i_{ln}$ for each location index $n$ of the whole volume and each label $l$ of the $N_{label}$ labels. The vote maps of the same label are averaged across all the atlases to generate the 15-channel output of the weighted-voting-subnet [$S^{init} = \{S^{init}_l, l = 1,2, \ldots, N_{label}\}$ with voxel value $s^{init}_{ln}$], i.e. $s^{init}_{ln} = (\sum_{i=1}^{N_{atlas}} v^i_{ln})/N_{atlas}$ for each $n$ and $l$. Importantly, the average operation allows the network to take any number of atlases as inputs, i.e. the number of atlases in training can be different from those in the testing phase.

***Fine-tuning-subnet and atlas mask:*** The fine-tuning-subnet gives the network flexibility to adjust the output generated by the weighted-voting-subnet. It employs the same U-Net structure as the weighted-voting-subnet, with the only exception of having four levels of down/up sampling instead of three. It takes $S^{init}$ and the coordinate maps as inputs and generate output feature maps that are the same size of $S^{init}$ (15 channels for the 15 labels). Then, each channel of the feature map is multiplied by a mask generated by thresholding the mean votes of the corresponding label of all the atlases (threshold is set to 0.2 empirically). This is based on the assumption that final segmentation of each label should be inside the region of most of the atlas votes of that label. The final segmentation is generated by performing the *arg-max* operation across the 15 channels of the masked feature maps.

### 3.1.2 Implementation Details

*Obtaining the patch-level training and test sets.* Limited by the GPU memory capacity, we trained the proposed network using image patches. Each patch of the training set consists of patches of the target image and all the registered atlases (both images and candidate segmentations) at the same location. To obtain such a training set, we perform leave-one-out cross-validation among the training subjects, i.e. image and manual segmentation of one subject is left out as the target subject and the remaining subjects are treated as atlases in each experiment. To be noted, this cross-validation experiment is performed to train the network and it is different from the top-level 4-fold cross-validation mentioned in Sec. 2. Since each subject has bilateral MTL segmentations available, to maximize the number of atlases, we include the left-right flipped atlases in the training, i.e. each subject provides both left and right MTL as atlases. For each bilateral MTL, the atlases are registered to the target following these steps: (1) the atlases and target T2w MRI scans are cropped around the MTL using the preprocessing pipeline described in [14]; (2) the cropped T2w MRI scans are up-sampled to 0.4x0.4x0.4 mm$^3$; (3) T1w MRI is affinely registered and resampled to the cropped T2w MRI; (4) cropped atlases images are registered (multimodal deformable registration) to the cropped target image; (5) the atlases and the manual labels are warped to the space of the cropped target image. All the registrations are done using the Greedy registration toolbox (github.com/pyushkevich/greedy).

Twelve patches (10 centered on voxels with foreground and 2 on background) with size 72x72x72 voxels are sampled for the target image and all the registered atlases. T1w and T2w patches are normalized by subtracting the mean and dividing by the standard deviation. The same strategy was used to generate the patch-level test set by treating the test subject as the target and all the training subjects as the atlases.

*Augmentation strategies.* When generating the training samples, the most straightforward way is to use all the atlases and have their sequence fixed. However, this may result in a network that is not robust to different atlas combinations or different atlas sequences and, thus, will not generalize well in the test phase. This is important because the number of atlases in the test phase may be different from training (e.g. there will be one more atlas available in the test phase using the leave-one-out strategy). In order to overcome this limitation, when sampling each patch, we randomly selected (with replacement) $N_{atlas}$ out of all the available atlases ($N_{atlas}$ can be bigger than the number of available atlases). This sampling strategy may result in repeated presentation of some atlases, which is desired as it may teach the network to deal with correlated errors among the atlases, similar to the core idea of JLF [5]. In addition, random elastic deformation augmentation [15] is applied to all the training patches to double the size of the training set. In our experiment, random flipping and random rotation ($\leq 10°$) did not improve segmentation accuracy and thus was not performed in the interest of time.

*Other implementation details.* The model was implemented in PyTorch and trained using a NVidia Tesla P100 GPU available in the Google Cloud Platform using generalized Dice loss [16]. The model converged after 10 epochs when trained using Adam optimizer (initial learning rate was set to 0.0005 and was reduced by a factor of 0.2 every 2 epochs beginning at the 4$^{th}$ epoch). The batch size was set to 1, constrained by GPU memory. The deep-supervision scheme [17] (4 levels with weights of 1, 0.5, 0.2,

0.1) was adopted to in the fine-tuning-subnet to train DLF, which was found to be beneficial. In testing, we sampled the test images with dense Cartesian grid (36x36x36 voxels spacing) to generate patches, which were fed to the trained DLF network to generate the final segmentation.

### 3.2 Alternative Methods for Comparisons

#### 3.2.1 Conventional Label Fusion Methods

Conventional label fusion methods, including MV, SVWV and JLF, were performed together with a neighborhood search scheme [18] to obtain benchmark performance. In addition, we also performed corrective learning (CL) [19], which is commonly used together with JLF, to get the best performance of conventional MAS methods. A grid search was performed for each approach to determine the optimal set of hyper-parameters that yields the best generalized Dice similarity coefficient (GDSC) [20] of all the GM labels between the automatic and manual segmentations.

#### 3.2.2 Direct U-Net

A 3D U-Net [11] was trained to generate benchmark performance of direct DL method. Its architecture is almost the same as the fine-tuning-subnet. The U-Net was trained on patches of the multimodal images (20 foreground and 8 background patches per hemisphere per subject) sampled from the aligned cropped T1w and T2w MRI generated in Sec. 3.1.3. and augmented using the random elastic deformation augmentation [15]. The patch size was set to 72x72x72 voxels for fair comparisons. The U-Net was trained over 20 epochs using generalized Dice loss and Adam optimizer (initial learning rate was set to 0.0005 and was reduced by a factor of 0.2 every 4 epochs beginning at the $9^{th}$ epoch) with a batch size equal to 7. The dense-sampling approach described in Sec. 3.1.3 was used to generate segmentations of the test images. The same deep-supervision scheme was adopted to train the 3D U-Net.

We observed that the direct U-Net generates a lot of erroneous foreground labels at the edge of the cropped image, which could be due to training on patches. An additional post-processing step is required to generate reasonable final segmentation. This is done by multiplying the initial segmentation with a binary mask generated by taking the largest connected component of the foreground label. Interestingly, the proposed DLF framework does not have this issue which may be the benefit of the intrinsic spatial constraints of the MAS-based framework.

## 4 Experiments and Results

***Validation on the 3T MRI dataset.*** Segmentation accuracy is evaluated using the Dice similarity coefficient (DSC) (GDSC for GM labels together) between the automatic and manual segmentations. For the 3T dataset, we use the $1^{st}$ fold as the test set and the other 3 folds as the training set to search for the optimal sets of hyper-parameters that yield the best GDSC of all the GM labels for each method. Then, the same set of optimal parameters is applied in the other three cross-validation experiments (excluding the one using the first fold as the test set to reduce potential bias) to generate the evaluation results, reported in Table 1. Paired-sample t-test (2-sided) was performed between DLF and each of the other methods to evaluate the significance of the improvement. In addition, to visualize the locations where each method makes errors, we warped the errormaps (binary map with label 1 indicating disagreement between the automatic and manual segmentations) to an unbiased population template, built using the manual segmentations as in [21]. Mean errormaps of all the methods were generated by averaging the corresponding errormaps of all the test subjects, reported in Fig. 2.

**Table 1.** Mean (±standard deviation) DSC and generalized DSC (GDSC) of all subregions between automatic and manual segmentations in the test folds of the 3T MRI dataset. Volume of each label is provided for better interpretation DSC scores. For better interpretation, background color of each cell indicates the relative performance compared to the best (most red) and worst (most blue) performance in each row (the more red/blue, the closer to the best/worst performance respectively).

|        | Volume (cm$^3$) | MV | SVWV | JLF + CL | U-Net | DLF |
|--------|---|---|---|---|---|---|
| GDSC   | - | 70.8±*4.3** | 75.8±*3.8** | 77.9±*3.8** | 77.1±*4.8** | **80.1± *3.2*** |
| Hippo  | 2.84±*0.46* | 88.5±*2.1** | 91.5±*1.2** | 92.9±*1.3* | 91.5±*3.5** | **93.1±*1.1*** |
| CA1    | 0.67±*0.14* | 68.4±*6.0** | 73.3±*3.9** | 75.4±*4.2** | 77.6±*4.1* | **78.0±*2.8*** |
| CA2    | 0.07±*0.01* | 54.2±*9.6** | 61.3±*7.4** | 69.0±*5.1** | **72.3±*3.9*** | 72.2±*4.0* |
| CA3    | 0.16±*0.03* | 63.5±*5.2** | 68.1±*4.5** | 71.3±*3.5** | 73.7±*4.5** | **75.6±*4.2*** |
| DG     | 0.52±*0.11* | 75.1±*4.1** | 79.9±*2.8** | 82.1±*2.4** | 82.5±*3.1* | **83.3±*1.5*** |
| SUB    | 1.02±*0.20* | 75.2±*7.7** | 78.3±*7.1** | 83.1±*2.9** | 81.4±*5.4** | **84.0±*2.2*** |
| TAIL   | 0.40±*0.13* | 79.9±*3.2** | **81.8±*3.0*** | 80.1±*7.1* | 77.3±*8.3* | 78.8±*6.7* |
| ERC    | 0.87±*0.24* | 75.0±*3.9** | 78.6±*3.7** | 80.8±*3.7** | 80.3±*6.0** | **84.4±*3.3*** |
| BA35   | 0.63±*0.14* | 56.8±*10.1** | 64.4±*9.4** | 66.4±*8.9** | 69.4±*8.2** | **72.2±*6.9*** |
| BA36   | 1.74±*0.58* | 68.7±*6.4** | 76.3±*5.6** | 78.2±*5.3** | 74.6±*7.2** | **80.0±*6.0*** |
| PHC    | 0.60±*0.19* | 67.8±*9.1** | 71.2±*7.9** | 74.6±*8.0* | 75.5±*7.8* | **76.4±*6.1*** |

Note: *: p < 0.05 compared to DLF. Hyper-parameters: SVWV: β = 0.05; JLF+CL: β = 2.0. The optimal patch size is 3x3x1voxels for both SVWV and JLF+CL. The optimal search radius is 4x4x1 voxels for SVWV and 3x3x1 voxels for JLF+CL.

Overall, DLF significantly improves segmentation accuracy in most of the subregions compared to label fusion methods with the biggest improvement in CA1-3, ERC and BA35, which are the most important subregions in Alzheimer's disease research. On the other hand, U-Net performs significantly worse than DLF in majority of the subregions (except CA1-2, DG, TAIL and PHC) with whole hippocampus, ERC, BA35 and BA36 being the worst. Notably, judging from the difference in errormaps in Fig. 3 (top), the improvements of DLF over JLF+CL and U-Net are not the same. Smaller error is observed in anterior and posterior BA35 as well as the anterior TAIL boundary compared to JLF+CL (white arrows) while medial ERC and lateral BA36 are the hotspots compared to U-Net (yellow arrows). This suggests that DLF is able to integrate the differential strong aspects of MAS and DL to improve segmentation accuracy.

**Figure 3.** Mean errormap of DLF and anatomical labels of the 3T (top) and 7T (bottom) MRI datasets. Difference in mean errormaps between alternative methods and DLF are shown on the right with red or blue indicating the alternative methods having more or less mean errors respectively.

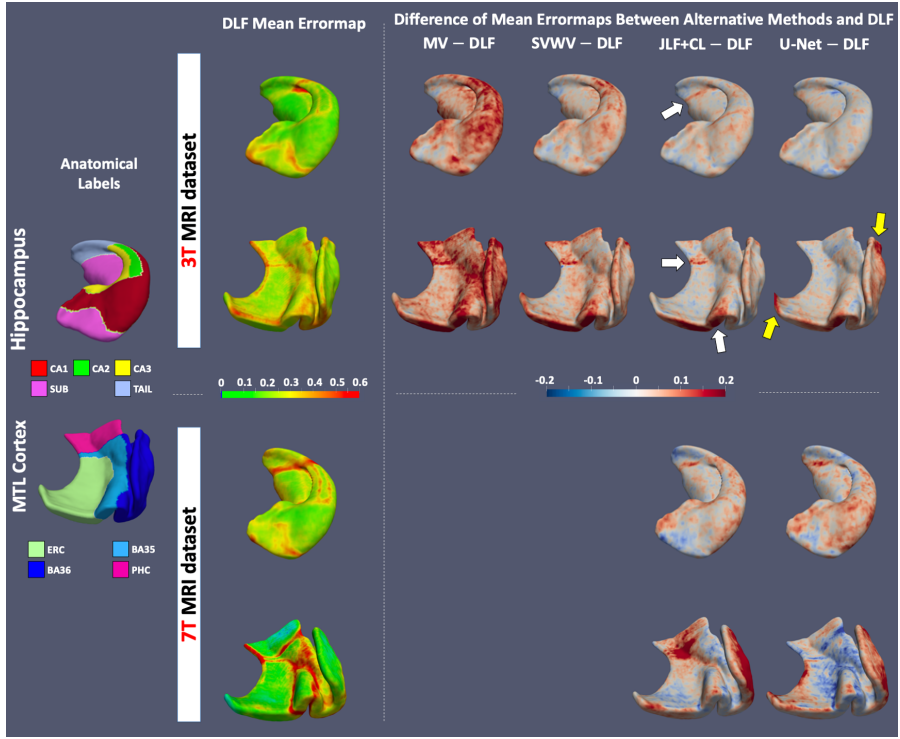

***Testing the generalizability on the 7T MRI dataset.*** To test whether the proposed DLF can generalize better to unseen data of a similar task, we directly apply DLF as well as JFL+CL (the best conventional label fusion method in Table 1) and U-Net that are trained on the 3T to the 7T dataset and evaluate the segmentation accuracy in terms of DSC/GDSC between the generated segmentations and the manual ones of the 7T dataset. As shown in Table 2 and Fig. 3 (bottom), DLF significantly outperforms JFL+CL and U-Net in most of the subregions (except for SUB and TAIL compared to JFL+CL and PHC compared to U-Net). In general, JFL+CL and U-Net perform better in segmenting hippocampal subfields (CA1-2, DG, SUB, TAIL) and MTL cordial subregions (BA35, BA36, PHC) respectively. Interestingly, DLF is able to take advantage of both approaches and generalizes the best in the unseen 7T dataset. One thing to be noted is that the performance in the 7T dataset is worse compared to the 3T dataset in terms of smaller DSC scores, which is expected given that the manual segmentations were generated by different raters on images with different contrast. Indeed, the segmentation errors (bottom second column in Fig. 3) are mainly concentrated at tissue boundaries between neighborhood GM subregions, where the highest inter-rater variability is expected, rather than gray/white matter or gray matter/cerebral spinal fluid boundaries.

**Table 2.** Mean (±standard deviation) DSC and generalized DSC (GDSC) of all subregions between automatic and manual segmentations in the 7T MRI dataset. Volume of each label is provided for better interpretation of the DSC scores. Background color of each cell follow the same rule in Table 1.

|  | Volume (cm$^3$) | JLF+CL | U-Net | DLF |
|---|---|---|---|---|
| **GDSC** | - | 66.7±*5.8** | 65.3±*7.6** | **70.7±*5.6*** |
| **Hippo** | 2.74±0.44 | 89.0±*2.1** | 80.2±*7.7** | **89.4±*2.2*** |
| **CA1** | 0.64±0.16 | 71.5±*4.1** | 61.6±*9.9** | **74.2±*5.6*** |
| **CA2** | 0.06±0.01 | 62.1±*9.4** | 58.0±*15.5** | **65.3±*10.7*** |
| **CA3** | 0.13±0.03 | 58.7±*7.9** | 58.9±*11.8** | **64.6±*7.6*** |
| **DG** | 0.47±0.09 | 77.7±*4.4** | 71.3±*10.6** | **79.2±*4.9*** |
| **SUB** | 0.97±0.14 | 78.4±*3.2* | 68.1±*7.8** | **78.7±*4.6*** |
| **TAIL** | 0.46±0.17 | 71.4±*5.8* | 67.1±*8.0** | **72.0±*5.5*** |
| **ERC** | 0.72±0.17 | 69.1±*9.3** | 68.3±*12.5** | **73.8±*8.2*** |
| **BA35** | 0.53±0.09 | 52.5±*10.3** | 60.0±*8.5** | **62.3±*8.8*** |
| **BA36** | 1.44±0.42 | 56.7±*10.5** | 60.8±*10.0** | **64.6±*8.3*** |
| **PHC** | 0.53±0.17 | 61.1±*13.7** | **70.5±*8.9*** | 67.7±*10.0* |

Note: *: p < 0.05 compared to DLF. Hyper-parameters are the same as that in Table 1.

**Table 3.** Mean (±standard deviation) DSC and generalized DSC (GDSC) of all subregions between automatic and manual segmentations in the training fold (the first fold) of the 3T MRI dataset with components of DLF taken out. No statistical test was performed because of the limited sample size. Note: w/o = without. Background color of each cell follow the same rule in Table 1.

|  | w/o fine-tuning-subnet | w/o atlas mask | w/o weighted-voting-subnet | DLF |
|---|---|---|---|---|
| **GDSC** | 72.9±*6.0* | 77.6±*4.2* | 75.7±*4.5* | **78.1±*4.7*** |
| **Hippo** | 88.2±*3.8* | 93.4±*1.2* | 93.3±*1.4* | **93.6±*1.3*** |
| **CA1** | 66.0±*8.1* | 74.6±*4.8* | **76.0±*4.8*** | 75.6±*5.2* |
| **CA2** | 55.3±*15.5* | 71.2±*5.9* | **72.9±*2.4*** | 71.6±*4.7* |
| **CA3** | 66.9±*7.5* | 74.0±*5.6* | **73.5±*4.9*** | 73.4±*6.7* |
| **DG** | 81.7±*2.6* | 81.8±*2.5* | 82.2±*2.3* | **83.2±*2.8*** |
| **SUB** | 79.7±*3.1* | 81.6±*3.2* | 81.8±*3.4* | **82.4±*3.2*** |
| **TAIL** | 71.5±*9.8* | 79.3±*3.6* | 78.7±*5.8* | **79.4±*4.2*** |
| **ERC** | 77.9±*4.3* | **84.2±*2.8*** | 81.7±*4.3* | 83.7±*3.3* |
| **BA35** | 66.0±*12.6* | 71.8±*7.7* | 69.7±*8.1* | **72.0±*9.5*** |
| **BA36** | 70.6±*10.8* | 77.1±*5.3* | 71.1±*7.7* | **77.4±*5.7*** |
| **PHC** | 69.7±*10.7* | 69.0±*9.6* | 67.9±*8.8* | **70.4±*10.3*** |

***Evaluate the contributions of sub-components of DLF.*** To demonstrate the importance of the contribution of the important or unique sub-components of DLF, i.e. weighted-voting-subnet, fine-tunning-subnet and atlas mask, we also report in Table 3 the performance of DLF in the training fold (the first fold, rather than the other three folds as in Table 1 and Table 2) with each of them taken out. The results show that the fine-tuning-subnet contributes most (+6.2 in GDSC). Although the atlas mask brings small improvement (+0.5 in GDSC), it may be important in constraining spatial location of the final segmentation, potentially contributing to better generalizability.

## 5 Conclusion

In this work, we proposed an end-to-end 3D hybrid multi-atlas segmentation and deep learning segmentation pipeline. Experimental results on MTL subregion segmentation in the 3T dataset demonstrate significant improvement compared to both conventional multi-atlas segmentation approaches and U-Net, a direct DL method. Further, results on the unseen 7T dataset highlight the better generalizability of the proposed DLF approach. Future work includes investigating the relative contribution of T1w and T2w MRI, evaluating the proposed pipeline in other tissue segmentation tasks, speeding up registration using learning-based methods, and performing more comprehensive comparisons with other learning-based label fusion approaches. [1]

[1] This work was supported by NIH (grant numbers R01-AG056014, R01-AG040271, P30-AG010124, R01-EB017255, R01-AG055005) and Google Cloud.